\documentclass[11pt,a4paper]{article}
\usepackage{amsmath,amssymb,amsthm,amsfonts}
\usepackage[dvips]{graphicx}
\usepackage{epsfig}
\usepackage{siunitx}	
\usepackage{subfigure}
\usepackage{float}
\usepackage{verbatim}
\usepackage[font=small]{caption}
\usepackage{soul}

\usepackage{color}

\title{Nanocrystal growth via the precipitation method}
\author{C. Fanelli\footnote{Centre de Recerca Matem\`{a}tica, Campus de Bellaterra  Edifici C, 08193 Bellaterra, Barcelona, Spain.} \footnote{Departament de Matem\`{a}tica Aplicada, Universitat Polit\`{e}cnica de Catalunya, Barcelona, Spain.} \footnote{Barcelona Graduate School of Mathematics, Campus de Bellaterra,  Edifici C, 08193 Bellaterra, Barcelona, Spain}, V. Cregan\footnote{MACSI/PMTC, University of Limerick, Ireland.}, F. Font\footnotemark[1], T.G. Myers\footnotemark[1] \footnotemark[2].}

\def\a{\alpha}
\def\b{\beta}

\def\d{\delta}

\def\f{\frac}

\def\p{\partial}
\def\q{\quad}

\def\tx{\text}

\def\({\text{\huge (}}
\def\){\text{\huge )}}

\def\]{\text{\huge ]}}
\def\[{\text{\huge [}}

\def\tx{\text}

\providecommand{\keywords}[1]{\textbf{\textit{Keywords:}} #1}

\newcommand{\Oc}{\mathcal{O}}
\newcommand{\refb}[1] {(\ref{#1})}		 	   
\newcommand{\Fig}[1] {Figure \ref{#1}}	   
\newcommand{\ud}{\mathop{}\!\mathrm{d}}

\newcommand{\et}{\emph{et al}.\ }
\newcommand{\bi}{\begin{itemize}}
\newcommand{\ei}{\end{itemize}}
\newcommand{\be}{\begin{equation}}
\newcommand{\ee}{\end{equation}}
\newcommand{\ba}{\begin{align}}
\newcommand{\ea}{\end{align}}

\newcommand\nc{\newcommand}
\nc\pad[2]{\frac{\p #1}{\p #2}} \nc\padd[2]{\frac{\p^2 #1}{\p
{#2}^2}} \nc\nd[2]{\frac{d #1}{d #2}} \nc\pat[2]{\frac{D #1}{D
#2}} \nc\ov{\overline} \nc\degree{^{\circ}} \nc\ord[1]{{\cal
O}(#1)} \nc\ra{\rightarrow} \nc\Ra{\Rightarrow} \nc\dint{{\mbox ~
d}}

\DeclareMathOperator{\Da}{Da}		

\begin{document}
\maketitle

\begin{abstract}

A mathematical model to describe the growth of an arbitrarily large number of nanocrystals from solution is presented.
First, the model for a single particle is developed. By non-dimensionalising the system we are able to determine the dominant terms and reduce it to the standard pseudo-steady approximation. The range of applicability and further reductions are discussed. An approximate analytical solution is also presented. The one particle model is then generalised to $N$ well dispersed particles. By setting
$N=2$ we are able to investigate in detail the process of Ostwald ripening. The various models, the $N$ particle, single particle and the analytical solution are compared against experimental data, all showing excellent agreement. By allowing $N$ to increase we show that the single particle model may be considered as representing the average radius of a system with a large number of particles. Following a similar argument the $N=2$ model could describe an initially bimodal distribution. The mathematical solution clearly shows the effect of problem parameters on the growth process and, significantly, that there is a single controlling group. The model provides a simple way to understand nanocrystal growth and hence to guide and optimise the process.

\end{abstract}


\section{Introduction}
\label{Sect:Introduction}

Nanoparticles (NPs) are small units of matter with dimensions in the range 1-\SI{100}{nm}. They exhibit many advantageous, size-dependent properties
such as magnetic, electrical, chemical and optical, which are not observed at the microscale or larger \cite{murray2000synthesis, grieve2000,bell2003, tanabe2007}. Consequently the ability to produce monodisperse particles that lie within a controlled size distribution is critical.

There exist a number of NP synthesis methods, including gas phase and solution based synthesis techniques. Although the first method can produce large quantities of nanoparticles, it produces undesired agglomeration and nonuniformity in particle size and shape. Precipitation of NPs from solution avoids these problems and is one of the most widely used synthesis methods \cite{mantzaris2005liquid}. The typical strategy is to cause a short nucleation burst in order to create a large number of nuclei in a short space of time, and the seeds generated are used for the latter particle growth stage. Thus the temporal separation of nucleation and growth occurs, as proposed by La Mer and Dinegar \cite{mer1952nucleation, lamer1950theory}, is applied.  The resulting  system  consists of varying sized particles.
Small NPs are more unstable than larger ones and tend to grow or dissolve faster. Thus at relatively high
monomer concentrations size focussing occurs (leading to monodispersity). When the monomer concentration is
depleted by the growth some smaller NPs shrink and eventually disappear while larger particles continue to grow, thus leading to a broadening of the size distribution (Ostwald ripening).

The particle size distribution (PSD) can be refocused by changing the reaction kinetics. For example, Peng \et \cite{peng2001mechanisms} observed size focusing during Cadmium Selenide growth following the injection of additional solute.  Bast\'us \et \cite{bastus2011kinetically, bastus2014synthesis} were also able to induce size focusing of gold and silver nanoparticles by the addition of extra solute and adjusting the temperature and pH. This type of technique for size focussing is still rather \emph{ad hoc} in that the precise relationships between particle growth, system conditions and the final PSD are not fully understood \cite{schmid2008clusters}. Hence, in practice, the optimal reaction conditions are usually ascertained empirically or intuitively.

In the 1960's Lifshitz and Slyozov \cite{lifshitz1961kinetics} and, independently, Wagner \cite{wagner1961theorie} were amongst the first to provide theoretical descriptions of Ostwald ripening. Their classical theory,  hereafter referred to as LSW theory,  consisted of a system of three coupled equations: a growth equation for a single particle, a continuity equation for the PSD and a mass conservation expression for the concentration.  They solved the model to obtain pseudo-steady-state asymptotic solutions for the average particle radius and  PSD.  Lifshitz and Slyozov \cite{lifshitz1961kinetics} focused on diffusion-limited growth, where growth is limited by the diffusion of reactants to the particle surface, while Wagner \cite{wagner1961theorie} considered growth limited by the reactions at the particle surface (i.e. reaction-limited growth).
In fact recent work described by Myers and Fanelli  \cite{myersfanelli2018} have shown that, within the restrictions of the steady-state assumption, the model cannot distinguish between diffusion or reaction driven growth, so both approaches are equally valid. For this reasons authors using either mechanism, or both, have been equally successful in approximating experimental data.

Experimental studies on NP growth \cite{chuang2008, pan2006, su2010cdsegrowth} show that LSW theory may provide good predictions for  the particle size but the observed PSDs were typically broader and more symmetric. Possible explanations for this disparity is that LSW theory does not account for the finite volume of the coarsening phase $\phi$, and that it assumes a particle's growth rate is independent of its surroundings.  In addition, LSW theory does not indicate how long it takes to reach the final state. A further issue is that it purports to describe the dynamics in the initial stages of the growth process. In \cite{myersfanelli2018} it is proven that the pseudo-steady solution does not hold for small times.

Many studies  have  modified and built on the pioneering analysis of LSW theory. Ardell \cite{ardell1972effect} and Sarian and Weart \cite{sarian1966kinetics}  extended LSW theory to systems for non-zero $\phi$ (i.e.  the mean distance between particles is finite).
Several authors \cite{brailsford1979dependence,voorhees1984solution1, voorhees1984solution2} have addressed the  shortcomings of LSW theory by statistically averaging the diffusional interaction of a particle of a given size with its surroundings to demonstrate that the resulting PSD becomes broader and more symmetric with increasing $\phi$.
The inclusion of stochastic effects, due to temperature and changes in concentration, in the modified population balance model of Ludwig \et \cite{ludwig1994influence} led to broader PSDs in line with experimental data. The population balance approach of Iggland and Mazzotti \cite{iggland2012population} was used to examine the evolution of non-spherical particles at the beginning of growth.

Most of the above studies were in relation to micron or larger-sized particles. As measurement techniques have advanced many researchers have applied LSW theory and the related modifications to the study of nanoparticle growth.
Talapin \et \cite{talapin2001evolution} used a Monte Carlo approach to simulate the evolution of a nanoparticle PSD subject to diffusion-limited growth, reaction-limited growth and mixed diffusion-reaction growth.  In contrast to other treatments, their simulations gave PSDs narrower than those predicted by LSW theory. This was explained by the fact that they considered much smaller particles. Their main conclusion was that Ostwald ripening occurs much more rapidly for nanoparticles while PSDs are narrower than in their microscale counterparts.
Similarly, Mantzaris \cite{mantzaris2005liquid} used a population balance formulation and a moving boundary algorithm  to study the diffusion and reaction-limited growth regimes.

Another  issue which is particularly relevant in the context of nanoparticles is the applicability of the Ostwald-Freundlich condition (OFC) which relates the radius of the particle, $r_p^*$, to its solubility, $s^*$. This condition can be written as
\be
s^* = s_\infty^* \exp{\left( \frac{2 \sigma V_M}{r_p^* R_G T} \right)}  \equiv  s_\infty^* \exp{\left( \frac{\alpha}{r_p^*} \right)} \,,
\label{eqn:OstwaldFreundlich}
\ee
where  $s_\infty^*$ is the solubility of the bulk material, $\sigma$ the interfacial energy, $R_G$ the universal gas constant, $T$ the absolute temperature. The capillary length $\alpha = 2 \sigma V_M /(R_G T)$ defines the length scale below which curvature-induced solubility is significant \cite{talapin2001evolution}. This equation shows that the particle solubility increases as the size decreases (which promotes Ostwald ripening).  One approximation  to the OFC is to assume  that the exponential term in \refb{eqn:OstwaldFreundlich} can be  linearised to give the two term expression $s^* \approx s_\infty^* (1 + \a/r_p^*) $\cite{lifshitz1961kinetics, liu2007ostwald, sugimoto1987preparation, wagner1961theorie}. Obviously this expansion, which is based on $\a/r_p^*$, is invalid for nanoparticles where capillary length is of the same order of magnitude as the particle radius \cite{myersfanelli2018}.
Mantzaris \cite{mantzaris2005liquid} used an expansion for the exponential term in the OFC with $n$ terms and showed  that increasing $n$ led to higher average growth rates and a narrowing of the PSD. However, when comparing his simulation to experimental data for CdSe nanoparticles from \cite{peng1998kinetics}, he applied the linearised version for the solubility.
Talapin \et \cite{talapin2001evolution}, noting that for nanoparticles of the order 1-\SI{5}{nm} the linearised OFC may be incorrect, applied the full condition.

In the following we  begin by analysing the growth of a single particle. This is the basic building block for more complex models. The treatment leads to equations similar to those of standard LSW theory, however we arrive at them following a non-dimensionalisation which highlights dominant terms and those which may be formally neglected. In this way we can ascertain which standard assumptions are appropriate and, more importantly, which are not. Under conditions which appear easily satisfied for nanocrystal growth the governing ordinary differential equation has an explicit solution, in the form $r_p=r_p(t)$ and also shows that the growth is controlled by a single parameter which may be calculated by comparison with experiment. This section closely follows the work described in \cite{myersfanelli2018}.
The single particle model is obviously incapable of reproducing Ostwald ripening, where larger particles grow at the expense of smaller ones. Consequently we then generalise the model to deal with a large number of particles. In the results section we compare the analytical solution with that of a full numerical solution and experimental data for the growth of a single particle and show excellent agreement between all three. By setting the number of particles to two in the general model we are able to clearly demonstrate Ostwald ripening. Simulations with $N=$10 and 1000 particles demonstrates that increasing $N$ leads to increasingly good agreement between the average radius and that predicted by the single particle model. The single particle model may thus be considered as a viable method for predicting the evolution of the average radius of a  group of particles.

\section{Growth of a single particle}
\label{Sect:SingleParticle}

\begin{figure}[ht]
\centering
\includegraphics[width=0.8\textwidth]{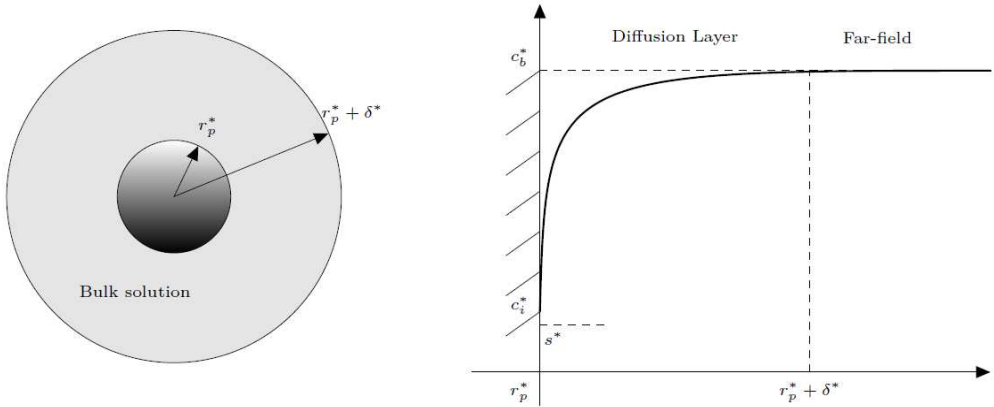}
\caption{Schematic of a single nanoparticle with radius $r_p^*$ and the surrounding monomer concentration profile where $s^*$, $c_i^*$ and $c_b^*$ are the particle solubility, the concentration at the surface of the particle and the far-field concentration, respectively.}
\label{Fig1}
\end{figure}

As shown in \Fig{Fig1}, we initially focus on a single, spherical nanoparticle, with radius $r_p^*$ in a system of particles. The $^*$ notation represents  dimensional quantities. The assumption is that particles are separated at large but finite distances compared to their radius. Their morphologies remain nearly spherical and particle aggregation is neglected. Thus, the mass flow from each particle can be represented as a monopole source located at the center of the particle \cite{voorhees1992} and the problem becomes radially symmetric.
We assume  the standard La Mer model \cite{mer1952nucleation}, such that there has been a short nucleation burst and the system is now in the period of growth.

The monomer concentration, $c^*$, is described by the classical diffusion equation in spherical coordinates
\be
\frac{\partial c^*}{\partial t^*} = \frac{D}{r^{*2}}  \frac{\partial^*}{\partial r^*}\left(r^{*2} \frac{\partial c^*}{\partial r^*} \right) \, .
\label{ConcentrationEquation}
\ee

This holds in the diffusion layer $[r_p^* , r_p^*+ \delta^*]$ where $r^*$ is distance from the centre of the particle, $t^*$ is time and $D$ is the constant diffusion coefficient.  To conform with standard literature (see for example \cite{mantzaris2005liquid, sugimoto1987preparation, viswanatha2007growth}), we have included a diffusion layer of length $\delta^*$ around the particle, where the concentration adjusts from the value at the particle surface to the value in the far-field. Equation (\ref{ConcentrationEquation}) is subject to

\be
c^*(r_p^*,t^*)=c_i^*(t^*) \,, \qquad \,
c^*(r_p^*+ \delta^*,t^*)=c_b^*(t^*) \,, \qquad \,
c^*(r^*,0)=c_{b,0}^*  \q  \tx{for $r^*>r_p^*$} ~\,,
\label{BulkConcentration}
\ee

where $c^*_i$ is the concentration adjacent to the particle surface, $c_b^*$ is the concentration in the far-field and $c_{b,0}^*$ is a constant describing the initial concentration when the solution is well-mixed.
The monomer concentration in the far-field $c_b^*(t^*)$ will be derived via mass conservation. The value at the particle surface $c_i^*$ is very difficult to measure \cite{sugimoto1987preparation}, hence it is standard to work in terms of the particle solubility.

The particle solubility $s^*$ (with the same dimensions as concentration) is given by the Ostwald--Freundlich condition \refb{eqn:OstwaldFreundlich}. If  $s^* <c_b^*$ then  monomer molecules diffuse from the bulk towards the particle to  react with the  surface and the particle grows, whereas if $s^* > c_b^*$ the particle shrinks.

In order to determinate an expression for the concentration at the particle surface, we consider two equivalent expressions for the mass flux at the particle surface, $J$. Firstly, Fick's first law states that the flux of monomers passing through a spherical surface of radius $r^*$ is

\be
J= 4 \pi r^{*2} D\frac{\partial c^*}{\partial r^*} \, .
\label{eqn:FluxExpression1}
\ee

At the surface of the sphere the flux must also follow a standard first order reaction equation

\be
J=4 \pi r^{*2} k (c_i^*-s^*) \, ,
\label{eqn:FluxExpression2}
\ee

where $k$ is the reaction rate, which is assumed to be constant for both growth and dissolution contributions. Equating  \refb{eqn:FluxExpression1} with \refb{eqn:FluxExpression2}  gives

\be
c_i^*=s^* +  \frac{D}{k} \frac{\partial c^*}{\partial r^*} \bigg|_{r^*=r_p^*} \, ,
\label{eqn:ciExpression}
\ee

which defines the concentration $c_i^*$ for the surface condition of (\ref{BulkConcentration}).

To complete the boundary conditions in the system, we require an expression for the time-dependent bulk concentration, $c_b^*(t^*)$. The mass of monomer in the system is constant because the bulk material is assumed well-mixed during the entire process.
Mass conservation of the monomer atoms in the particle and surrounding  solution is then

\be
\frac{1}{N_0} M_p c^*_0 = M_p c^*_b(t) \left[ \frac{1}{N_0} - \frac{4 \pi}{3} {r_p^*}^3 \right] + \frac{4 \pi}{3}  \rho_p {r_p^*}^3
\label{eqn:MassConservationFarField}
\ee

where $\rho_p$ is density, $M_p$ is molar mass and $N_0$ the population density. Since particle occupy a small region in comparison to their radius $4 \pi N_0 {r_p^*}^3/3  \ll 1$, also the molar volume $V_M=M_p/\rho_p$. Equation (\ref{eqn:MassConservationFarField}) then leads to

\be
c^*_b(t) \approx c^*_0 - \frac{4 \pi N_0}{3V_M} {r_p^*}^3.
\label{eqn:MassConservationFarFieldRewritten}
\ee

which will be used for the far-field concentration in (\ref{BulkConcentration}).

As stated earlier, the diffusion equation must be solved on a domain $r^* > r_p^*$, where the particle radius is an unknown function of time. The flux of monomer to the particle is responsible for the particle growth

\begin{equation}
V_M J = \frac{d}{dt^*} \left( \frac{4}{3} \pi {r_p^*}^3 \right)= 4 \pi {r_p^*}^2 \frac{d r_p^*}{dt^*} \, .
\label{eqn:FluxRadiusEquation}
\end{equation}

Eliminating $J$ between (\ref{eqn:FluxRadiusEquation}) and (\ref{eqn:FluxExpression1}) yields

\be
\frac{\ud r_p^*}{\ud t^*} = V_M D \frac{\partial c^*}{\partial r^*}\bigg|_{r^*=r_p^*}  \, .
\label{ParticleGrowthEquation}
\ee

This is subject to the initial condition $r_p^*(0) = r_{p,0}^*$, where $r_{p,0}^*$ is the initial particle radius.

The governing system is now fully defined and consists of equation \refb{ConcentrationEquation}, subject to the initial and boundary conditions \refb{BulkConcentration}, where $c^*_i$ is defined by \refb{eqn:ciExpression} and $c_b^*$ by \refb{eqn:MassConservationFarFieldRewritten}, and the unknown particle radius satisfies \refb{ParticleGrowthEquation}. As there is no analytical solution to the system, we proceed to simplify the problem and use numerical approximations in order to understand the behaviour of the solution.


\subsection{Nondimensionalisation}
\label{Nondimensionalisation}

A complex mathematical model can often be reduced to a simpler form by estimating the relative magnitude of terms. A standard way to achieve this is by writing the system in non-dimensional form. Here the  model  is nondimensionalised via
\be
r =\frac{r^*}{r_{p,0}^*}\, , \q\q
r_p =\frac{r_p^*}{r_{p,0}^*}\, ,   \q\q
t =\frac{t^*}{\tau^*} \, ,  \q\q
c =\frac{c^*-s^*_0}{\Delta c} \, ,\q\q
s =\frac{s^*-s^*_0}{\Delta c} \, ,
\label{eqn:DimensionlessVariables}
\ee
where $\Delta c=c^*_{b,0}-s^*_0$ represents the driving force for particle growth and  $s^*_0 = s_\infty^*\exp{(\alpha/r_{p,0}^*)}$ is the initial particle solubility.
The concentration and growth equations yield two possible time scales $\tau^*_D \sim r_{p,0}^{*2}/D$ and  $\tau^*_R \sim r_{p,0}^{*2}/(V_M D c_0)$, respectively.
To focus on  particle growth we choose the growth time scale $\tau^*=\tau^*_R$ and the governing system is now transformed to

\begin{align}
& \varepsilon \frac{\partial c}{\partial t}  = \frac{1}{r^2}  \frac{\partial}{\partial r}\left(r^2 \frac{\partial c}{\partial r} \right) \, ,
\label{DimensionlessEquationDiff} \\
& \frac{\ud r_p}{\ud t}  = \frac{\partial c}{\partial r}\bigg|_{r=r_p}  \, ,
\label{DimensionlessEquationStef} \\
& c(r_p,t) = s  +  \Da \frac{\partial c}{\partial r} \bigg|_{r=r_p} \,, \qquad
c(r_p+\d,t) = \bar{c}_0 - \beta r_p^3, \, , \qquad
\label{DimensionlessBConditions} \\
& c(r,0) =1  \, , \qquad
r_p(0)=1 \, ,
\label{DimensionlessIConditions}
\end{align}
where
\be
s= s_\infty  \exp{(\omega/r_p)} -s_0  \, ,
\label{eqn:ce_definition}
\ee
and
\be
\varepsilon = V_M \Delta c \,,  \qquad
\d =\f{\delta^*}{r_{p,0}^*}\, ,   \q\q\q
\Da=\f{D}{k r_{p,0}^*}  \,, \qquad
\omega=\frac{\alpha}{r_{p,0}^*} \,,\qquad
\b = \f{4 \pi N_0 r_{p,0}^{*3}}{3 V_M}  \,, \qquad
\bar{c}_0 = \frac{c^*_0-s^*_0}{\Delta c} \,.
\label{eqn:Parameters}
\ee

The above system contains a number of nondimensional groups. The first, $\varepsilon$, is generally very small for nanoparticle growth. For example, Peng \et \cite{peng1998kinetics} studied Cadmium Selenide nanoparticles, with a capillary  length of 6nm and initial radii in the range  $1-100\,$nm, so that $\varepsilon = \Oc(10^{-3})$. In general it should be expected that $\varepsilon \ll 1$. If we look at the time scales, we see that $\tau^*_D/ \tau^*_G = V_M \Delta c = \varepsilon \ll 1$. Physically, this indicates that growth is orders of magnitude slower than the diffusion time scale, that is, the concentration adjusts much faster than growth occurs and so the system can be considered as pseudo-steady. In terms of the mathematical model, this means that the time derivative can be omitted from the concentration equation, but since time also enters into the problem through the definitions of $r_p$  and $c_b$ this is a pseudo-steady-state situation rather than a true steady-state.

The parameter $\Da$ is an inverse Damk\"{o}hler number  measuring the relative magnitude of diffusion to surface reactions  \cite{mantzaris2005liquid}. In the past similar models have been simplified by  considering  diffusion-limited growth ($\Da\ll1)$ or surface reaction limited growth ($\Da \gg 1$).
In practice both mechanisms play a role. In \cite{myersfanelli2018} it is shown that the diffusion limited case requires either $k \to 0$, which results in zero growth, or the concentration adjacent to the particle surface matches the solubility $c_i \simeq s^*$ throughout the process. Similary the reaction driven growth requires $D \to 0$, which again indicates no growth, or $c_i \simeq c_b$ throughout the process. Therefore, we will place no restrictions on $\Da$.

A common simplification is to assume $\omega \ll 1$ which reduces the OFC, \eqref{eqn:OstwaldFreundlich}, to a constant $s^* = s_{\infty}^*$ or a linear approximation is used, see \cite{liu2007ostwald, sugimoto1987preparation}. This significantly simplifies the analysis. However, for particles that have just nucleated or very small nanoparticles $\omega$ is not small and the simplification is not appropriate.
Despite the large errors in the prediction of $s^*$ caused by the small $\omega$ assumption authors obtain good matches to data. In \cite{myersfanelli2018} it is shown that this is because the pseudo-steady model is not valid for early times when the particle is small. By the time the model is valid so is the linearisation. Basically, the variation of $s^*$ plays a minor role in the study of the growth of a single nanoparticle. However, this is not the case with multiple particles where Ostwald ripening is driven by the delicate balance between the bulk concentration and the particle solubility.

The reason why the pseudo-steady model is invalid at small times is due to the thickness of the boundary layer $\delta(t)$. The model involves the assumption $\delta(t) \gg r_p$ yet initially, when the fluid is well-mixed $\delta(0)=0$. Only when the boundary layer is sufficiently thick is it reasonable to apply the pseudo-steady model. In \cite{myersfanelli2018} it is shown through comparison with experiments that the initial stage can last for the order of 100s. The shift to the pseudo-steady model can often be identified simply by looking at the trend in the data. In the following we will present the model with the full OFC and then an approximation where it is neglected. We will also neglect early data points when matching to experimental data.


\subsection{Pseudo-steady state solution}
\label{Sect:PerturbationSolution}

Since  $\varepsilon = \Oc(10^{-3})$ and all variables have been scaled to be $\Oc(1)$ then neglecting terms of order $\varepsilon$ should result in errors of the order 0.1\%. Consequently, we neglect the time derivative in the diffusion equation and obtain the pseudo-steady state form
\be
 \frac{1}{r^2}  \frac{\partial}{\partial r}\left(r^2 \frac{\partial c}{\partial r} \right) = 0 \, .
\label{eqn:PseudoSteadyStateCEquation}
\ee
After integrating and applying the boundary conditions we obtain
\be
c = -\f{A}{r} + B \, ,
\label{eqn:PseudoEquationC}
\ee
where
\be
A=\f{r_p^2 (r_p+\d)(c_b-s)}{r_p \d + \Da(r_p +\d)} \,, \qquad
B= s+ A\left( \f{1}{r_p} + \f{\Da}{r_p^2} \right).
\label{eqn:FullConstantConcentration}
\ee
There is no way to calculate $\delta(t)$ in the pseudo-steady approach. A time-dependent treatment, such as that described in \cite{myers2010optimal} is required. Hence the standard method is to assume $r_p\ll\d$, which reduces \refb{eqn:FullConstantConcentration} to
\be
A=\f{r_p^2 (c_b-s)}{r_p + \Da} \,, \qquad
B= c_b \,.
\label{eqn:FullConstantConcentrationReduced}
\ee
Substituting \refb{eqn:PseudoEquationC} into the Stefan condition \refb{DimensionlessEquationStef}  leads to
\be
\frac{\ud r_p}{\ud t} = \frac{c_b-s}{\Da+r_p} =\frac{\bar{c}_0 - \beta r_p^3 +  s_0 - s_\infty\exp(\omega/r_p)}{\Da+r_p} \, .
\label{eqn:PseudoEquationRp}
\ee
Hence, the problem has been reduced to the solution of a single first-order ordinary differential equation for $r_p$.
It is a highly nonlinear equation which must be solved numerically. The assumption that $r_p \ll \delta$ means it only holds for relatively large times. Approximate solutions, in various limits, may be found in the literature. For example if we take $c_b$ constant and $\omega$ sufficiently small for the linear approximation to the exponential to hold then equation (\ref{eqn:PseudoEquationRp}) may be integrated in the limits of large and small $\Da$.
In \cite{myersfanelli2018} it is shown that for sufficiently large times, for a single particle, the variation of $e^{\omega/r_p}$ does not affect the solution in which case equation (\ref{eqn:PseudoEquationRp}) may be integrated analytically to find an implicit solution of the form $t=t(r)$. By identifying negligible terms they are able to invert this to find an explicit solution, $r=r(t)$ which depends on a single parameter,
\begin{align}\label{myersfanelli}
r_p = \frac{r_m}{2}\frac{\left[1 +2 f(r_{p0})\exp\left(\frac{t-t_0}{G}\right)-\sqrt{-3+12f(r_{p0})\exp\left(\frac{t-t_0}{G}\right)}\right]}{\left[-1+f(r_{p0})\exp\left(\frac{t-t_0}{G}\right)\right]}\,,
\end{align}
where $r_m$ is the experimental maximum radius, $t_0$ is the time at which the second growth stage is judged to have begun $r_{p0}$ the radius at this time and $f(r_{p0}) = (r_m^2+r_mr_{p0}+r_{p0}^2)/(r_m-r_{p0})^2$. If $t_0$ is greater than the true value, this should not affect the results. This is discussed in further detail in \cite{myersfanelli2018}.  The unknown parameter $G$ is defined as
\be
\label{GDef}
G = \frac{1}{6ab} \frac{ak+bD}{akbD} ~,  \qquad a^3 = V_m (c_0^*-c^*_{eq}) ~ ,\quad b^3 = \frac{4}{3} \pi N_0 ~ .
\ee
Its value is obtained by comparison with experimental data. Once $G$ is determined, then the diffusion coefficient ($D$), the reaction rate ($k$), the solubility of the bulk material ($s_{\infty}$) and population density ($N_0$) may be systematically retrieved. In \cite{myersfanelli2018} it is stated that $ak\approx bD$, hence $G \approx 1/(3a^2 b k) = 1/(3ab^2 D)$. Further, since $c_0^* \gg c_{eq}^*$ a reasonable approximation is $a^3 = V_m c_0^*$. Growth stops when the maximum radius $r_m^* = a/b$ is achieved.


\section{Evolution of a system of $N$ particles}
\label{Sect:SystemNParticles}

We now  extend the single particle model to an arbitrarily large system of particles. The particle radii, initial radii and solubilities  are denoted $r^*_{i}$, $r^*_{i,0}$ and $s_i^*$, respectively, where $i$  represents the $i^{\text{th}}$ particle and $i=1\ldots N$.
We nondimensionalise via \refb{eqn:DimensionlessVariables} with the only difference being that the mean value of the radii $\bar{r}_{p,0}^*$ replaces
the length scale $r_{p,0}^*$.
It has to be noted that this also affects the concentration scale, $\Delta c$, through the initial solubility. Hence in what follows, all dimensionless parameters are the same as those defined in \refb{eqn:Parameters}, except with $r_{p,0}^*$ and $s_0^*$ replaced by $\bar{r}_{p,0}^*$ and $\bar{s}_0^*$, respectively.

Under the pseudo-steady approximation and assuming that there are no interparticle diffusional interactions, the growth of each particle is now described by an equation of the form \eqref{eqn:PseudoEquationRp} with $r_{i}$ and $s_{i,0}$ substituting for $r_p$ and $s_0$, respectively.
The bulk concentration equation must account for all particles, that is

\begin{equation}
\frac{1}{N_0} M_p c^*_0 = M_p c^*_b(t^*) \left[ \frac{1}{N_0} - \frac{4 \pi}{3} \sum_{i=1}^N{r_i^*}^3 \right]+ \frac{4 \pi \rho_p}{3} \sum_{i=1}^N{r_i^*}^3
\end{equation}
where $N$ may decrease with time due to Ostwald ripening. Assuming that the solution is sufficiently dilute that $4 \pi N_0/3 \sum\limits_{i=1}^N r_{i}^{*3} \ll 1$ and also $V_M = M_p/\rho_p$, we obtain
\be
c_b^*(t^*) \approx c^*_0 - \frac{4 \pi N_0}{3 V_M} \sum\limits_{i=1}^N r_{i}^{*3} .
\label{DimensionalMassConservationNParticles}
\ee
In dimensionless form the problem is then governed by the system of differential equations
\be
\frac{\ud r_{i}}{\ud t} = \frac{c_b  - s_i}{\Da +r_{i}} =\frac{\bar{c}_0 - \bar{\beta} \sum\nolimits_{i=1}^{N} r_{i}^3 + \bar{s}_0- s_\infty\exp(\omega/r_{i})}{\Da+r_{i}} \,
\label{eqn:PseudoEquationRpNParticles}
\ee
for each $i = 1, \ldots , N$.
Equation \eqref{eqn:PseudoEquationRpNParticles} represents a system of $N$ non-linear ODEs which must be solved numerically.


\section{Comparison of model with experiment}
\label{Sect:Parameter}

The accuracy of the various forms of the mathematical model will now be ascertained through comparison with the experiments on CdSe nanocrystal synthesis reported by Peng {\it et. al.} \cite{peng1998kinetics}.
Certain parameter values concerning the experiment and CdSe are provided
in that paper, others, such as $D, k, s_{\infty}^*, N_0$ must be inferred. Here they will be determined through fitting to equation \eqref{myersfanelli}. Since this only contains one free parameter the fitting is a very simple process. We then show that the pseudo-steady state model (\textit{PSS model}) gives virtually identical results to equation \eqref{myersfanelli}. Once it has been established that the analytical solution and the PSS model give such good correspondence and match to experimental data we move on to comparing with the numerical results.
The PSS is an approximation to the full system defined by equations (\ref{DimensionlessEquationDiff}-\ref{eqn:ce_definition}), the analytical solution equation \eqref{myersfanelli} is an approximation to the PSS. Using the full system in the $N$ particle model would be extremely computationally expensive, for this reason
the PSS model is the basic component of the $N$ particle model. Consequently we demonstrate that the PSS closely matches the numerical solution of the full model (and consequently so does equation \eqref{myersfanelli}). The $N$ particle model is then examined. First, by setting $N=2$ we are able to demonstrate Ostwald ripening. We go on to show that as $N$ increases the prediction for the average radius tends to the analytical solution for a single particle as $N$ becomes large.

\subsection{Parameter estimation via the analytical solution}

\begin{figure}[ht]
	\centering
	\includegraphics[width=0.5\textwidth]{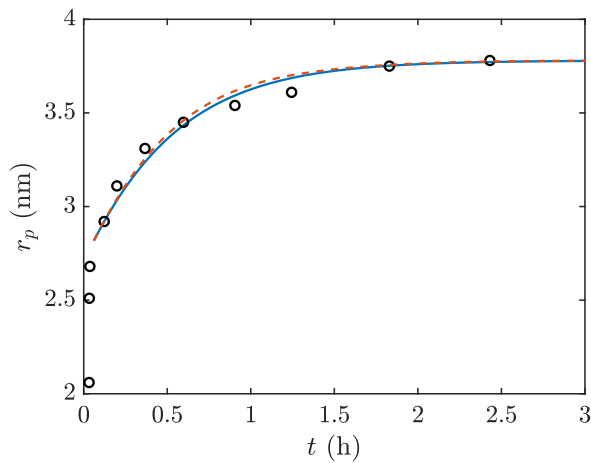}
	\caption{The circles represent the experimental data from Peng \et \cite{peng1998kinetics} and the solid line the corresponding least-squares fit to the analytical solution, equation \eqref{myersfanelli} with $G=958$, of Myers \& Fanelli \cite{myersfanelli2018}. The dashed line is the PSS model with the same value of $G$.}
	\label{Fig:NParticleGrowthVSPeng}
\end{figure}

In Figure \ref{Fig:NParticleGrowthVSPeng} we show the first eleven data points from \cite{peng1998kinetics}. As discussed earlier not all data points correspond to the pseudo-steady regime, here it is clear that the first three points follow a linear trend so these will be neglected. In the experiment extra monomer was added after three hours, so we have ignored all data beyond the eleventh point. Using the remaining eight data points in the nonlinear least-squares Matlab solver \textsf{lsqcurvefit} to fit to equation \eqref{myersfanelli} we obtain $G \approx 958$. The results of equation \eqref{myersfanelli}, with $G=958$, is shown as the solid line in Figure \ref{Fig:NParticleGrowthVSPeng}.

To determine the necessary parameters for the other models we first note that the maximum radius attained during this part of the experiment is $r_m^* \approx 3.8 $nm$=a/b = D/k$. The experimental concentration at the end of the growth process is known, this defines $c_{eq}^* = s_{\infty}^* e^{\alpha/r_m^*}$. This is enough information to determine $D, k, s_{\infty}^*, N_0$. The values taken from
\cite{peng1998kinetics} are shown as the first ten rows of  Table \ref{Table:PhysicalParameters}, the final four (in italics) are the ones calculated after $G$ has been determined. These values may be used for the pseudo-steady state (PSS) model, or the arbitrary $N$ model. The dashed line in Figure \ref{Fig:NParticleGrowthVSPeng} represents the result predicted by the PSS model. Clearly there is excellent agreement between this and the analytical solution, thus verifying the claim that the solubility may be set to a constant without greatly affecting the solution (provided the early time data is neglected).

\begin{table}[ht]
	\centering
	\begin{tabular}{ l  c  c  c }
		\hline
		Quantity & Symbol & Value & Units \\  \hline    \\
		Universal gas constant  & $R_G$ & 8.31 & \SI{}{J.mol^{-1}.K^{-1}}  \\ 
		Density & $\rho_p$ & 5816 & \SI{}{kg.m^{-3}} \\
		Molar volume & $V_M$ &  $3.29\times10^{-5}$  & \SI{}{m^{3}.mol^{-1}}  \\
		Molar mass & $M_p$ & 0.19 & \SI{}{kg.mol^{-1}} \\ 
		Solution temperature & $T$ & 573.15 & \SI{}{K} \\
		Surface energy  & $\sigma$ & 0.44  & \SI{}{J.m^{-2}} \\
		Capillary length & $\alpha$ & $6.00\times10^{-9}$ & \SI{}{m} \\
		Initial bulk concentration & $c_0^*$ &  55.33& \SI{}{mol.m^{-3}} \\
		Volume of the liquid & $V$ & $7.21 \times 10^{-6}$  & \SI{}{m^{3}} \\
		Maximum particle radius & $r_m$ & $3.78 \times 10^{-9}$  & \SI{}{m} \\
		\emph{Diffusion coefficient} & $D$ & $3.01 \times 10^{-18}$ & \SI{}{m^2.s^{-1}}\\
		\emph{Reaction rate}  & $k$ &  $7.97 \times 10^{-10}$ & \SI{}{m.s^{-1}}  \\
		\emph{Solubility of bulk material} & $s^*_{\infty}$ &  $5.53 \times10^{-2}$ & \SI{}{mol.m^{-3}} \\
		\emph{Population density}  & $N_0$ & $8.04 \times 10^{21}$ & \SI{}{No. m^{-3}}  \\
		\\ \hline
	\end{tabular}
	\caption{Physical parameters for the cadmium selenide (CdSe) nanoparticle synthesis method used by Peng \et \cite{peng1998kinetics}. The parameters in italics are not given in explicitly and are thus obtained via a fitting approach.  }
	\label{Table:PhysicalParameters}
\end{table}

\subsection{Validating the pseudo-steady state approximation}
\label{Subsect:pssapprox}

The PSS model  is described by equations \eqref{eqn:PseudoEquationC}-\eqref{eqn:PseudoEquationRp}. Since this forms the basis for the $N$ particle model it is important to verify its accuracy. We do this by comparison with the numerical solution of the full system \eqref{DimensionlessEquationDiff}-\eqref{DimensionlessIConditions} (referred to as the \textit{full model}). Although we have already shown that the PSS is very well approximated by the analytical solution and that for a single crystal the solubility variation may be neglected, we must employ the PSS in the $N$ particle model. This is because when an individual particle's solubility drops below the bulk concentration Ostwald ripening occurs. The analytical solution neglects variation in solubility, so cannot capture this behaviour.

Problems similar to the full model frequently occur in studies of phase change where it is termed the one-phase Stefan problem (one-phase because the temperature is neglected in one of the phases, this is analogous to neglecting the concentration in the crystal). Examples of one-phase problems occur in laser melting and ablation, Leidenfrost evaporation of a droplet and in supercooled materials. At the nanoscale there are many studies on nanoparticle melting and growth, see  \cite{FONT2017237,font2018,Myers09,Myers12}. The nanoparticle studies are particularly relevant, since they deal with a spherical geometry and at the nanoscale the melt temperature varies in a manner similar to the variation of the solubility in the current problem. For this reason we follow the numerical scheme outlined in these studies. It requires a standard boundary immobilization transformation and then a semi-implicit finite difference scheme is applied to the resulting equations. For further details see \cite{FONT2017237,font2018}. The PSS model requires the solution of a single nonlinear ordinary differential equation, \eqref{eqn:PseudoEquationRp}. To do this we  simply use the Matlab ODE solver \textsf{ode15s}. Once $r_p$ is determined the concentration is given by equation  \eqref{eqn:FullConstantConcentrationReduced}.

\begin{figure}[ht]
    \centering
    \subfigure[]{\includegraphics[width=0.45\textwidth]{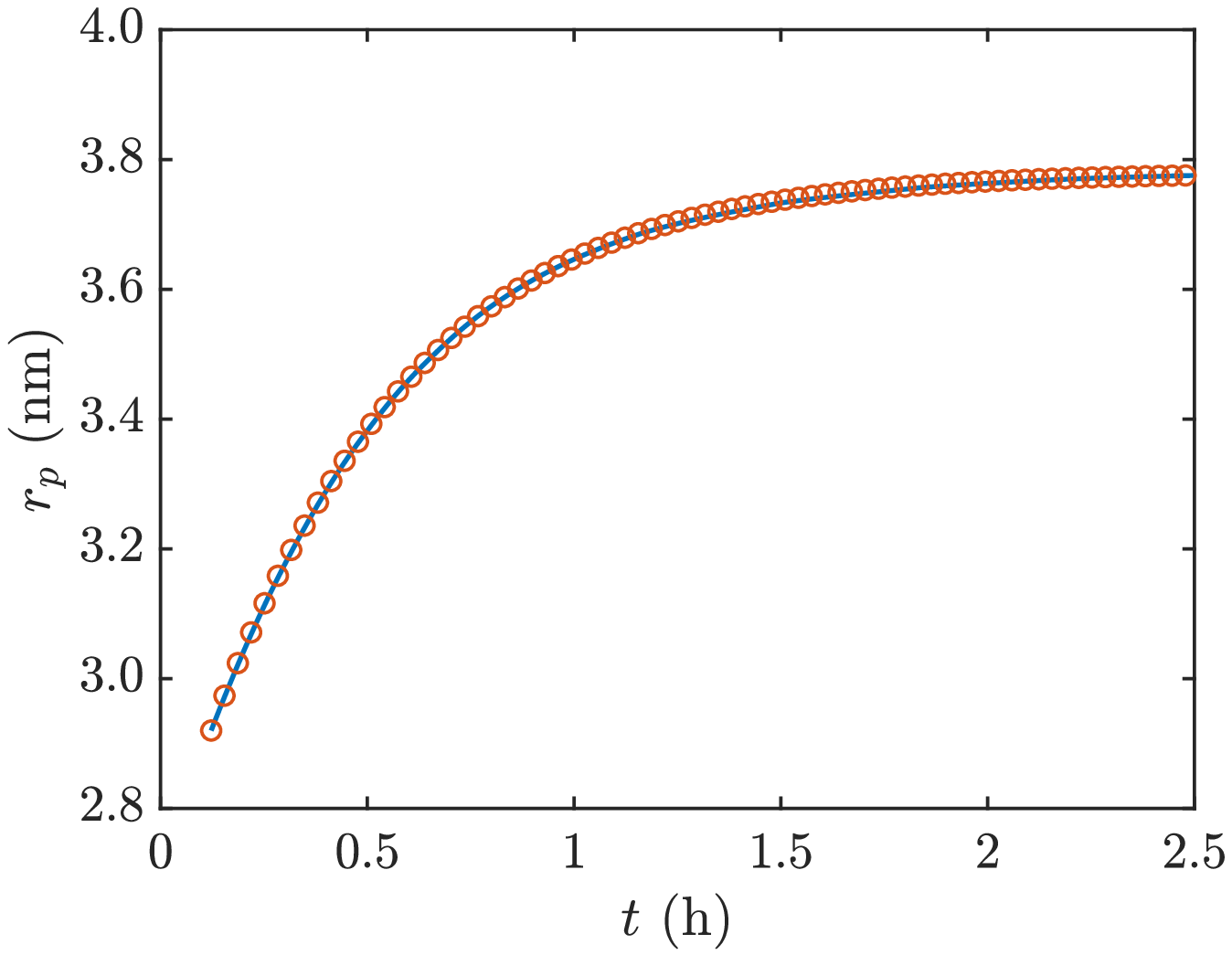}}
    \subfigure[]{\includegraphics[width=0.45\textwidth]{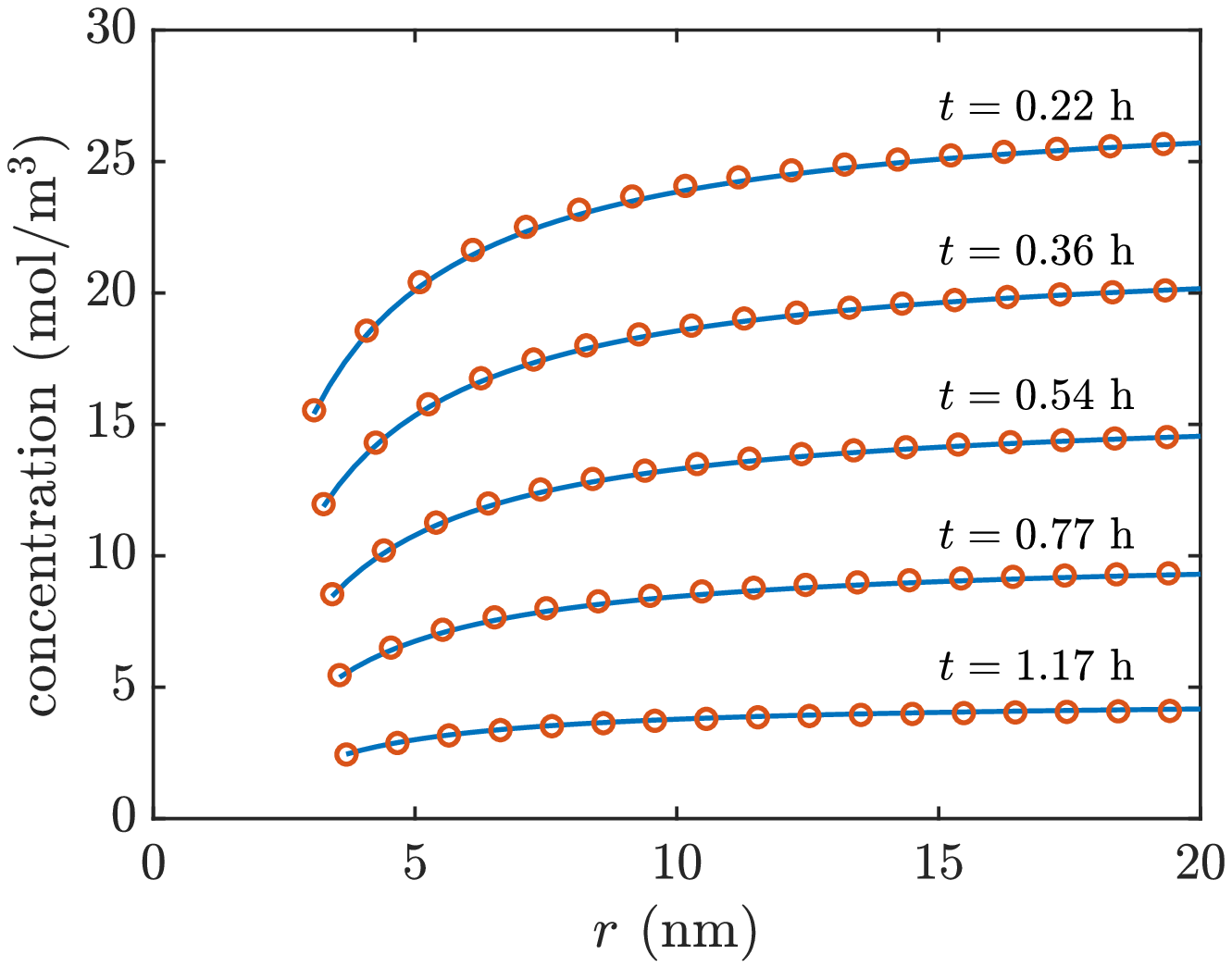}}
	\caption{Solution of the full and the PSS models (represented by circles and by a solid line, respectively) for the growth of a single particle. Panel (a) shows the evolution of the particle radius and panel (b) the concentration of monomer around the particle at five different times.  }
	\label{Fig:radiusandcon}
\end{figure}

In Figure \ref{Fig:radiusandcon} we compare the numerical solution of the full and PSS models using the parameters of Table \ref{Table:PhysicalParameters}, where panel (a) shows the evolution of the particle radius and panel (b) the concentration profile at five different times. In both cases the agreement between the full and PSS models is excellent, thus justifying the use of the simpler PSS model in the $N$ particle system.
Panel (a) shows how the particle grows rapidly until around $t \approx 1$ hr when the growth rate decreases, subsequently  the radius slowly approaches the maximum value of $r_p \simeq 3.8$ nm. This behaviour can be understood by analysing the concentration profiles presented in panel (b). The growth rate is proportional to the concentration gradient, see equation \eqref{ParticleGrowthEquation}. From Figure \ref{Fig:radiusandcon} (b) it is clear that the concentration gradient between the particle surface and the far field is relatively large at small times, leading to rapid growth. After $t \approx 1$ hr the concentration profile is practically flat, leading to a slow growth rate.


\subsection{Ostwald ripening with $N=2$}
\label{Subsect:2P}

Ostwald ripening (OR) occurs when the bulk concentration falls below a given particle's solubility. With a single particle the growth rate tends to zero as the solubility and bulk concentration become similar, hence OR will never occur. However, with a group of particles of different sizes theoretically OR must always occur. In practice this could take a very long time and so be difficult to observe. To demonstrate that the current model can predict OR we now investigate the simplest possible case, with two particles.

\begin{figure}[ht]
\centering
\subfigure[]{\includegraphics[width=0.45\textwidth]{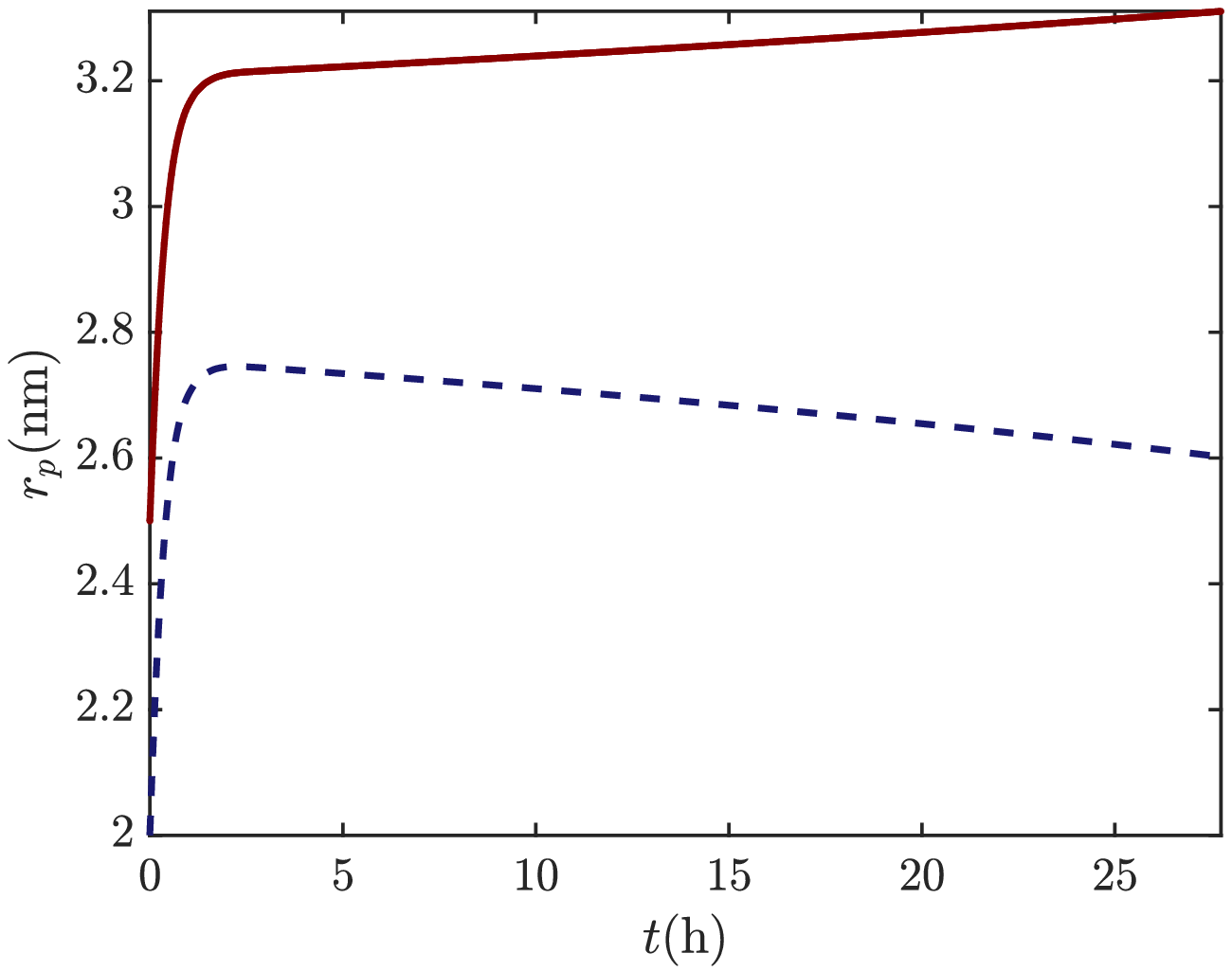}}
\subfigure[]{\includegraphics[width=0.45\textwidth]{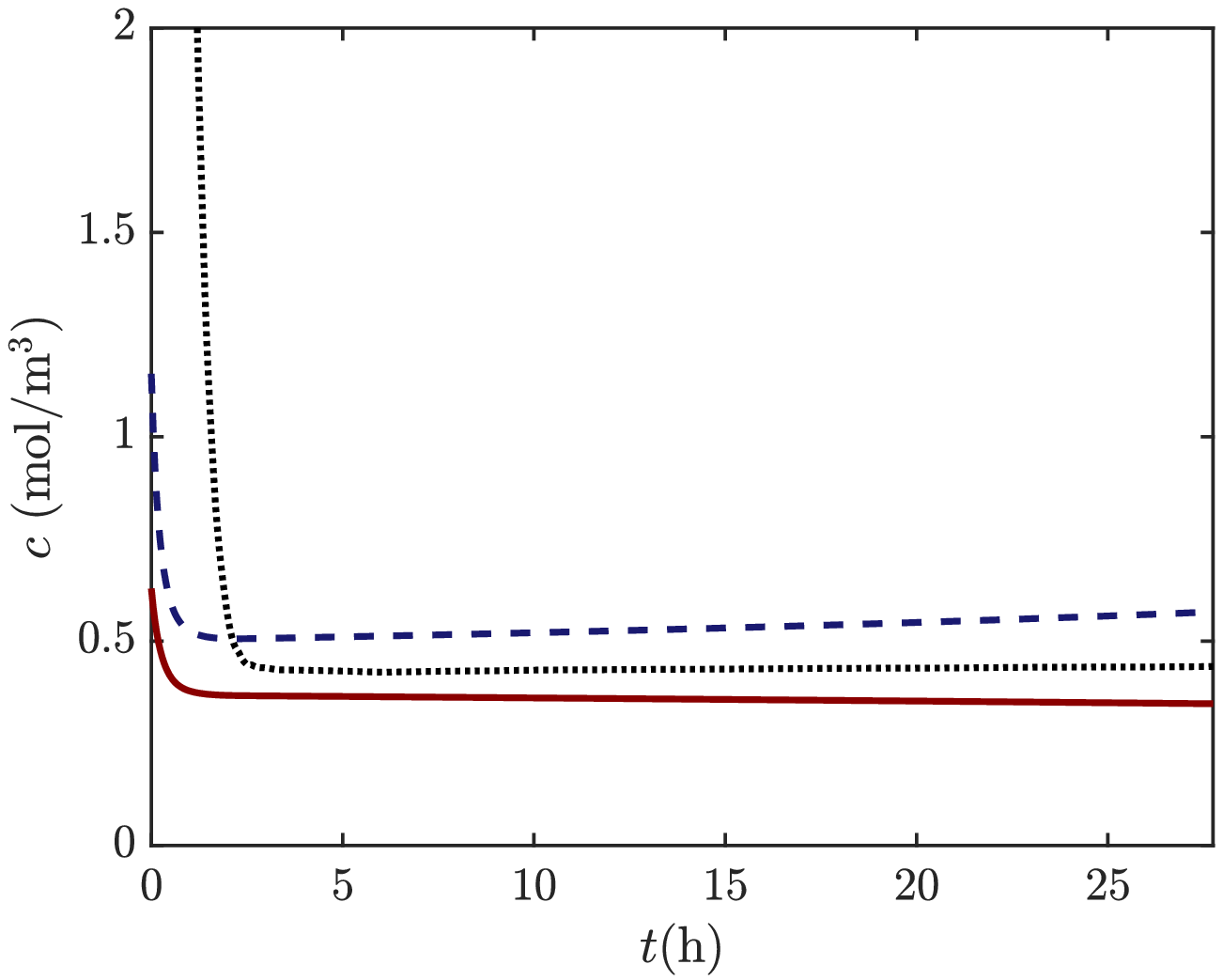}}
\caption{Evolution of two CdSe nanoparticles. (a) Change in time of the radii of two particles with initial radii of \SI{2}{nm} (dashed line) and \SI{2.5}{nm} (solid line). (b) Change in bulk concentration (dotted line) and solubilities of smaller (dashed line) and larger (solid line) particles.}
\label{Fig:TwoParticleGrowth}
\end{figure}

The system is  defined by equation \eqref{eqn:PseudoEquationRpNParticles} with $N=2$. We take parameter values from Table \ref{Table:PhysicalParameters} and choose initial radii \SI{2}{nm}  and \SI{2.5}{nm}. The governing equations may be solved again using the Matlab ODE solver \textsf{ode15s}. Results are presented in \Fig{Fig:TwoParticleGrowth}. The first figure shows the evolution of the radii for more than 25 hours. The solid line represents the evolution of the \SI{2.5}{nm} particle, the dashed line is the  \SI{2}{nm} one. As can be seen, for small times both particles grow rapidly however, after around 1.7 hours the smaller particle starts to shrink, while the larger one grows linearly.
In \Fig{Fig:TwoParticleGrowth}(b) the variation of the particle solubility and concentration is shown, solid and dashed lines correspond to the 2.5, 2nm particle's solubility respectively, while the dotted line is the bulk concentration. With reference to the variation of the radius it is clear that the rapid growth phase corresponds to a sharp decrease in the bulk concentration. Initially the solubility of each particle is below the bulk concentration and decreases as $r_p$ increases. Ostwald ripening begins when the solubility of the smaller particle crosses the $c_b$ curve, $c_b = s_1$ at $t \simeq 1.7$ hr, subsequently its size decreases. The solubility of the larger particle keeps slowly decreasing, in keeping with its slow growth, and remains below the bulk concentration until the end of the simulation. If we continued the simulation the smaller particle would eventually disappear.

\subsection{$N$ particles system}
\label{Subsect:NP}

To simulate the experiments of \cite{peng1998kinetics} we consider a distribution of $N$ nanoparticles where
the initial distribution is generated by random numbers, with an initial mean radius $\bar{r}^*_{p,0}$ of \SI{2.92}{nm} and a standard deviation of $\sigma_o =8.9$\%. In the numerical solution if a particle decreases below 2nm it is assumed to break up and all the monomer enters back into the bulk concentration.

\begin{figure}[ht]
\centering
\subfigure[]{\includegraphics[width=0.45\textwidth]{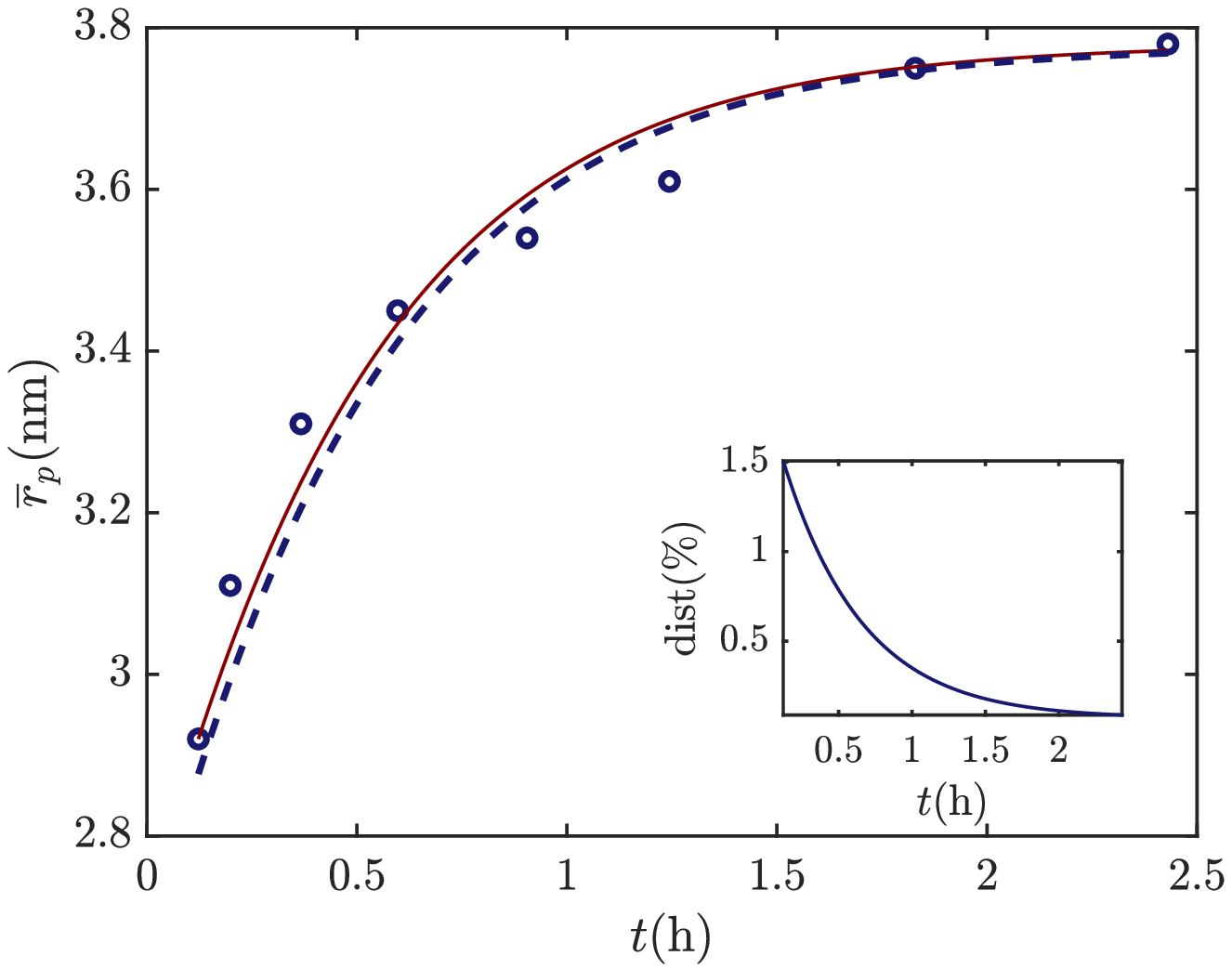}}
\subfigure[]{\includegraphics[width=0.45\textwidth]{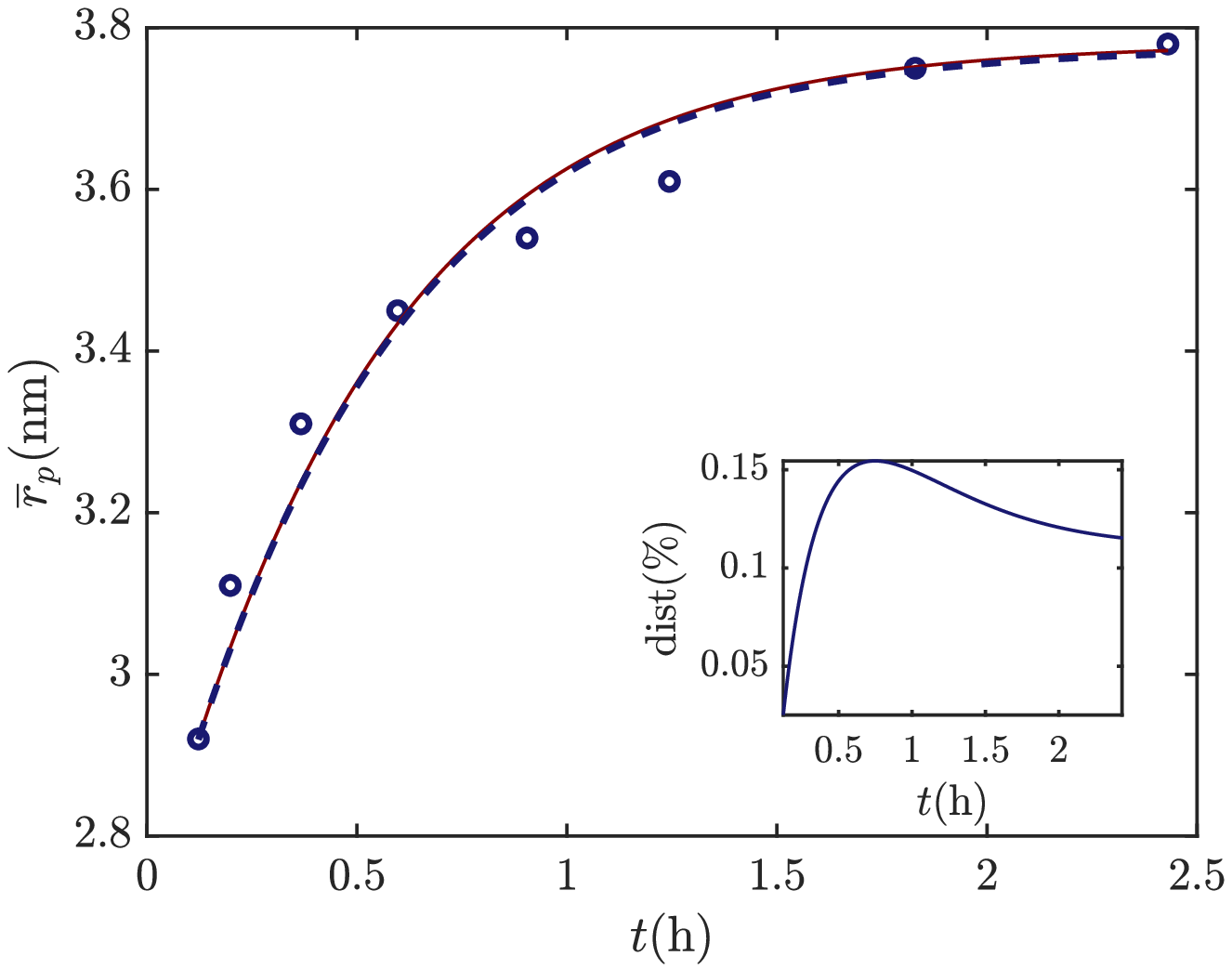}}
\caption{Comparison of the model for $N$ particles (dashed lines) with experimental data from Peng \textit{et al.}\cite{peng1998kinetics} (dots) using $N=10$ in (a) and $N=1000$ in (b). The solid lines represent the explicit solution for the one particle model, equation (\ref{myersfanelli}). The inset plots show the percentage difference between the models.}
\label{Fig:NParticleGrowthVSPeng_10&1000particles}
\end{figure}

In
\Fig{Fig:NParticleGrowthVSPeng_10&1000particles} we compare the prediction for the average radius of 10 and 1000 CdSe particles (dashed lines)
with the corresponding data of Peng \et \cite{peng1998kinetics}. The single particle analytical solution for $r(t)$, equation
(\ref{myersfanelli}), is shown as a solid line. The inset shows the difference between the $N$ particle and analytical solution. In
\Fig{Fig:NParticleGrowthVSPeng_10&1000particles}(a) the maximum difference between the two solutions is of the order 1.5\%, which decreases rapidly with time. The solution with $N=1000$, shown in 
\Fig{Fig:NParticleGrowthVSPeng_10&1000particles}(b), has a maximum difference of the order 0.15\% from the analytical solution.

In practice $N$ would be much higher. In Table \ref{Table:PhysicalParameters} the population density is given as $N_0 =8.04 \times 10^{21}$ crystals/m$^3$, so in a volume $V \approx 7 \times 10^{-6}$m$^3$ we would expect around $10^{16}$ crystals. The Figure demonstrates that as $N$ increases the solution tends to the analytical solution. Given that $N$ is typically very high it is then clearly not necessary to solve the large system. Obviously the analytical solution is easier to understand and implement than a $10^{16}$ particle model. However, it is important to note that in the present example there is no significant Ostwald ripening. From Figure \ref{twoparticlesradius_colors.eps} we observe defocussing starts around 1.7 hours and after nearly 30 hours the radius of the smaller particle has only decreased by 7\%. In the experimental data used here extra monomer is added to the solution after three hours and we stop our calculations then. So, in the absence of significant OR we may assume that the analytical solution may be used to predict the average evolution of nanocrystal growth. If OR is to be modelled an $N$ particle model
should be used, since this accounts for the solubility of each particle.


\section{Conclusions}
\label{Sect:Conclusions}

We have developed a model for the growth of a system of $N$ particles, where $N$ may be arbitrarily large. The model involves a system of first order nonlinear ordinary differential equations, which are easily solved using standard methods. The basis of the $N$ particle model is the pseudo-steady approximation presented in \cite{myersfanelli2018}. This incorporates the particle solubility variation which then permits the model to capture Ostwald ripening.

It has been shown that the pseudo-steady model for a single particle has an accurate approximate explicit solution. This was verified by comparison with the full pseudo-steady model. The explicit solution shows that there is a single main parameter controlling nanocrystal growth.

The main drawback to the single particle model is that it cannot capture Ostwald ripening, whereby smaller particles disappear at the expense of larger ones. By studying the system with $N=2$ we were able to emulate Ostwald ripening on a very simple system.
By allowing $N$ to become large and calculating the average particle radius we showed that the results approached the single particle explicit solution, which may thus be considered to represent the average growth of a large distribution of particles.
A consequence of this is that the $N=2$ model can equally well represent the average radii for an initially bimodal distribution of nanocrystals. An $N>2$ model can represent a much larger distribution of particles.

The main advantage of the current method is that since the single particle model may be solved analytically, and this accurately describes the average radius of a distribution, then the controlling parameters are apparent. This allows us to adjust them and so optimise the growth process, paving the way for efficient large scale production. Since we only have to deal with a single particle the numerical solution is rapid (almost instantaneous), as opposed to previous large scale, time-consuming calculations.


\section{Acknowledgements}

Claudia Fanelli acknowledges financial support from the Spanish Ministry of Economy and Competitiveness, through the “Maria de Maeztu” Programme for Units of Excellence in R\&D (MDM-2014-0445). Tim G. Myers acknowledges the support of Ministerio de Ciencia e Innovacion Grant No. MTM2017-82317-P. Francesc Font and Vincent Cregan acknowledge financial support from the Obra Social ``la Caixa'' through the programme Recerca en Matem\`{a}tica Col$\cdot$laborativa. The authors have been partially funded by the CERCA Programme of the Generalitat de Catalunya.


\bibliographystyle{plain}
\bibliography{Ostwald_bib}

\end{document}